\documentclass[journal]{IEEEtran}
\usepackage{amsmath}
\usepackage{algorithmic}
\usepackage{array}
\usepackage{graphicx}
\usepackage{cite} 
\usepackage{float}
\usepackage{xcolor}
\usepackage{amssymb}

\usepackage[breaklinks]{hyperref}

\begin{document}

\title{Multi parameter discrimination using multiple spectral troughs in a cascaded fiber sensor}


\author{Riming Xu, Yanbo Li, Xingnan Chen, Jin Wang*
\thanks{Riming Xu is with Physical Sciences and Engineering Division, King Abdullah University of Science and Technology, Thuwal 23955-6900, Saudi Arabia. Email: xu.rm@qq.com. 

Yanbo Li is with the Department of Civil and System Engineering, Johns Hopkins University, Baltimore 21218, USA.

Xingnan Chen is with the Collage of Mechanical and Electrical Engineering, Hainan Vocational University of Science and Technology, Haikou, Hainan, 571125, China.

Jin Wang is with Faculty of Science, Beijing University of Technology, Beijing 100124, China. Email: wangjinbjut321@163.com.
}}

\maketitle
\IEEEpeerreviewmaketitle

\begin{abstract}
Accurate monitoring of temperature, axial strain, and refractive index is critical for structural health monitoring, industrial process control, and environmental sensing. However, conventional optical fiber sensors are often limited by strong parameter cross sensitivity, poor discrimination capability, and increased system complexity when multiple sensing units are required. In this work, a compact multi-parameter optical fiber sensing platform is proposed based on a cascaded single-mode fiber, multimode fiber, and long-period fiber grating structure, combined with a wavelength-based spectral demodulation strategy.

Within the cascaded configuration, multiple characteristic spectral troughs arising from distinct physical mechanisms coexist in a single transmission spectrum. Interference-induced troughs are generated by the multimode fiber section, while a resonance-induced trough is introduced by the long-period fiber grating. Although none of these troughs responds exclusively to a single parameter, each exhibits simultaneous and linearly independent responses to temperature, axial strain, and refractive index with distinct sensitivity magnitudes and trends. Consequently, each trough can be described by a unique sensitivity vector, enabling robust multi-parameter discrimination through multi-wavelength spectral demodulation.

Systematic experiments demonstrate highly linear and repeatable wavelength shifts of three representative troughs under variations of temperature, strain, and refractive index. The measured temperature sensitivities are $-0.0733$, $-0.0833$, and $-0.0733$~nm/$^\circ$C, while the corresponding strain sensitivities reach $-0.6706$, $-0.6916$, and $-0.8026$~nm/$m\varepsilon$, all with coefficients of determination exceeding 0.99. Under refractive index modulation, two troughs show negligible responses below $0.0006$~nm/\%, whereas the third exhibits a pronounced sensitivity of $-0.1426$~nm/\% with excellent linearity.

By exploiting the differential sensitivity characteristics of multiple spectral troughs, effective discrimination of temperature, axial strain, and refractive index is achieved without additional sensing elements or complex optical paths. Long-term stability tests further confirm wavelength fluctuations within $\pm 0.01$~nm, highlighting the excellent stability and repeatability of the proposed sensing platform.
\end{abstract}


\section{Introduction}
\IEEEPARstart{O}{ptical} fiber sensors have been extensively investigated over the past decades due to their inherent advantages, including high sensitivity, immunity to electromagnetic interference, chemical inertness, compact size, and the capability of remote and distributed interrogation. These features make optical fiber sensing particularly attractive for applications in harsh or inaccessible environments, such as structural health monitoring, industrial process control, environmental monitoring, and biochemical analysis \cite{ramola2025comprehensive,gallareta2025advancements,zhang2025review,wang2022temperature}. With the continuous demand for higher measurement accuracy and system integration, increasing attention has been devoted to fiber based sensors capable of detecting multiple physical or chemical parameters within a single sensing platform \cite{anjana2025development,liu2025distributed,wang2023core,liu2025accurate}.

In practical environments, temperature, mechanical strain, refractive index, and surrounding refractive index often vary simultaneously and are strongly coupled. However, most fiber sensing mechanisms rely on spectral features whose wavelength shifts are inherently influenced by more than one external perturbation. For instance, wavelength shifts induced by temperature variations may be indistinguishable from those caused by axial strain or refractive index changes when only a single spectral feature is monitored \cite{zhang2025highly,choudhary2025polyphenylene}. Such cross sensitivity significantly limits the reliability of single parameter fiber sensors when deployed in complex environments, leading to systematic measurement errors and ambiguity in parameter interpretation \cite{singh2025surface,choudhary2025polyphenylene}. Consequently, multi parameter discrimination and crosstalk compensation have become central challenges in optical fiber sensing research \cite{redyuk2025compensation,song2025dual}.

A variety of strategies have been proposed to mitigate cross sensitivity in fiber optic sensing systems\cite{lyu2025fiber,davis2002application}. These strategies can be broadly categorized into structural decoupling approaches\cite{lin2025progress}, material or coating engineering\cite{qu2025advances,vendittozzi2025functional}, signal processing and data driven methods\cite{cao2025artificial}, and multi element integration schemes\cite{noori2025review,wang2025fiber}. 
Recent progress in deformation sensing for flexible robots has been systematically reviewed by Lin \textit{et al.}\cite{lin2025progress}, who summarized a wide range of structural decoupling strategies that exploit geometric design, cascaded sensing architectures, and mechanically distributed sensing units to discriminate complex deformation modes.
Structural decoupling methods typically rely on spatial separation or cascaded arrangements of different sensing elements to isolate individual perturbations, which often leads to increased device size and complex optical routing. 
Material based approaches have been extensively explored to improve sensing selectivity by introducing functional coatings or specialty materials onto the optical fiber surface. For example, Qu \textit{et al.} \cite{qu2025advances} systematically reviewed optical fiber sensors employing hygroscopic polymers, metal oxides, and functional nanocomposite coatings to selectively enhance refractive index and chemical sensitivity while suppressing cross sensitivity to temperature and strain.
More recently, data driven and algorithmic methods have been introduced for cross sensitivity mitigation in fiber optic sensing systems. Cao \textit{et al.} \cite{cao2025artificial} reviewed the integration of machine learning and artificial intelligence techniques with fiber Bragg gratings, interferometric sensors, and distributed fiber sensing platforms to achieve multi parameter demodulation through
data driven modeling and pattern recognition.

Among these strategies, one of the most widely adopted approaches is to integrate multiple sensing elements with distinct response characteristics into a single system and retrieve the target parameters through a sensitivity matrix inversion
process \cite{aidli2025dual}. Fiber Bragg gratings, long period fiber gratings, interferometric structures, and multimode interference devices have all been explored as building blocks for such systems
\cite{alhussein2025fiber,noori2025review,du2025structure,meyzia2025development}. Although this class of methods can significantly improve discrimination capability, many reported configurations suffer from increased system complexity, including multiple optical paths, parallel interrogation channels, or strict alignment requirements. These factors not only increase cost but also reduce scalability for practical deployment \cite{zhao2025applications}.

Cascade integration of different fiber structures along a single optical path offers an appealing solution to this problem. By allowing multiple sensing mechanisms to coexist spectrally within one transmission channel, cascade structures can significantly enhance system compactness while preserving multi parameter sensing capability . In particular, multimode interference structures formed by cascading single mode and multimode fibers can generate multiple stable spectral troughs with good repeatability and high signal to noise ratio \cite{wang2022temperature,liu2025accurate}. Meanwhile, long period fiber gratings introduce resonance troughs originating from core to cladding mode coupling, which are highly sensitive to environmental perturbations such as temperature, strain, and surrounding refractive index \cite{gizatulin2025using,deng2025scientometric}. The coexistence of interference induced troughs and resonance troughs within a cascaded structure provides rich spectral information for wavelength based demodulation.

Nevertheless, the presence of multiple spectral troughs does not imply that each
trough is exclusively sensitive to a single measurand. In practice, all spectral
troughs respond simultaneously to temperature, strain, and refractive index, albeit with
different sensitivity magnitudes and response trends governed by their formation
mechanisms \cite{wang2022temperature,berrocal2021crack,wang2022polarization,wang2022simultaneous}. Neglecting this intrinsic
cross sensitivity and assigning each trough to a single parameter may result in
misinterpretation and calibration instability. Therefore, multi parameter sensing
should be rigorously described by treating each spectral trough as a vector
response to multiple measurands rather than a scalar indicator of an individual
parameter.

In this work, we propose a compact and robust multi parameter discrimination strategy based on spectral troughs in a cascaded sensing configuration that integrates a single mode fiber section, a multimode fiber segment, and a long period fiber grating within a single optical path. The cascaded structure is deliberately designed so that multiple spectral troughs originating from different physical mechanisms coexist in distinct wavelength regions. The first two troughs arise from multimode interference within the multimode fiber segment, while the third trough originates from the resonance effect of the long period fiber grating. All troughs exhibit simultaneous responses to temperature, axial strain, and refractive index, but with clearly distinguishable sensitivity vectors.

Instead of attempting to isolate a single parameter sensitive feature, the proposed method employs wavelength based demodulation of multiple troughs combined with a matrix based sensitivity model. By constructing a sensitivity matrix from experimentally calibrated wavelength responses, the coupled effects of temperature, strain, and refractive index can be quantitatively decoupled through matrix inversion. This approach provides a unified and physically consistent framework for multi parameter discrimination using a single transmission spectrum, without the need for multiple interrogation channels or intensity based demodulation.

The contributions of this work can be summarized as follows. First, the spectral characteristics of the cascaded single mode fiber, multimode fiber, and long period fiber grating structure are systematically investigated under temperature, axial strain, and refractive index perturbations, demonstrating stable and highly linear wavelength responses of multiple spectral troughs. Second, a general matrix based demodulation model is established by treating each trough as a sensitivity vector with respect to multiple measurands, enabling rigorous crosstalk compensation. Third, the practical effectiveness of the proposed strategy is experimentally validated under strongly coupled conditions, including temperature disturbed concentration sensing and temperature strain coupled mechanical sensing, where conventional single parameter methods exhibit significant cross sensitivity.


\section{Experimental setup}

\subsection{Optical Fibers}

Standard single-mode fiber (SMF), with a core diameter of approximately $8~\mu\mathrm{m}$ and a cladding diameter of $125~\mu\mathrm{m}$, was used as the basic transmission medium and for cascaded interconnections between different sensing units. The SMF exhibits minimal transmission loss in the vicinity of $1550~\mathrm{nm}$, which matches the operating wavelength range of the broadband light source and the optical spectrum analyzer employed in this work.

Multimode fiber (MMF), with a core diameter of $110~\mu\mathrm{m}$ and a cladding diameter of $125~\mu\mathrm{m}$, was introduced to construct single-mode--multimode--single-mode (SMS) interference structures. The excitation and interference of multiple guided modes within the MMF segment enable effective multimode interference, which is utilized to enhance spectral modulation and sensing sensitivity.

Polarization-maintaining fiber (PMF) of the panda-type birefringent structure was employed as the primary sensing carrier. The PMF possesses stress-induced birefringence with well-defined fast and slow axes, providing stable polarization characteristics and high sensitivity to environmental perturbations such as temperature, strain, and refractive index variations. The PMF used in this study has a standard cladding diameter of $125~\mu\mathrm{m}$ and was purchased from ChangFei Company.

Long-period fiber gratings (LPFGs) were used as core sensing elements due to their strong coupling between core and cladding modes and their high sensitivity to external environmental changes. The LPFGs were fabricated with a grating period of approximately $450~\mu\mathrm{m}$ and a duty cycle of $1{:}1$. In the grating region, the protective polymer coating was partially removed to allow direct interaction between the surrounding environment and the evanescent field of the cladding modes, thereby enhancing sensitivity to temperature, strain, and ambient refractive index variations.

Prior to assembly, all optical fibers were thoroughly cleaned and their end faces were carefully prepared to minimize insertion loss, Fresnel reflections, and interference noise at fiber joints. This preprocessing ensured stable optical coupling and reliable spectral measurements throughout the experiments.

\subsection{Optical Source and Spectrum Measurement}
A broadband light source (BBS) was employed to provide a continuous spectral input, operating over a wavelength range of $600$--$1700~\mathrm{nm}$, which fully covers the working bands of all optical fiber sensors used in this study. The transmitted spectral signals were recorded using an optical spectrum analyzer (OSA, AQ6317B), featuring a measurement range of $800$--$1700~\mathrm{nm}$ and a maximum wavelength resolution of $0.01~\mathrm{nm}$. This high spectral resolution enables clear discrimination of the resonance dips of long-period fiber gratings (LPFGs) as well as fine interference fringes, thereby ensuring accurate detection of small wavelength shifts and meeting the requirements for high-precision multi-parameter demodulation.

\subsection{Temperature, Strain, and Refractive Index Control}
As schematically illustrated in Fig.~S1, dedicated experimental configurations were employed to independently regulate temperature, axial strain, and refractive index while maintaining a consistent optical interrogation scheme.
Temperature regulation was achieved using a temperature controlled chamber. During the measurements, the sensing region of the optical fiber was kept straight and firmly fixed to prevent additional strain induced by thermal expansion. Axial strain was applied using a uniformly distributed loading structure, ensuring that the optical fiber remained within the elastic deformation regime throughout the tensile process. Multiple measurements were performed at each strain level, and the averaged value was taken as the final result to improve measurement reliability.

Refractive index variations were introduced by preparing aqueous sodium chloride (NaCl) solutions with different concentrations. The LPFG sensing region was fully immersed in the solution, and the concentration was increased incrementally under constant-temperature conditions to eliminate the influence of temperature fluctuations on refractive index measurements.

\subsection{Fabrication of PMF Sagnac Loop}

To construct a temperature-reference sensing unit with high sensitivity and long-term stability, a Sagnac interferometric loop based on polarization-maintaining fiber (PMF) was employed in this work.

Under the present experimental conditions, the PMF segment in the Sagnac loop is rigidly fixed and free of external mechanical loading. Consequently, the axial strain applied to the fiber is negligible compared with the temperature-induced effect, and the wavelength shift of the interference spectrum is therefore dominated by temperature variations.

Owing to the intrinsic birefringence of PMF, orthogonally polarized modes propagating along the fast and slow axes experience different propagation constants. When light propagates in clockwise and counterclockwise directions within the Sagnac loop and recombines at the coupler, a stable phase difference is introduced, resulting in periodic interference fringes in the output spectrum.

In the PMF-based Sagnac interferometer, the incident light from the broadband source is equally split by a 3~dB fiber coupler into two beams, which propagate along the PMF loop in clockwise and counterclockwise directions, respectively. Owing to the intrinsic birefringence of the polarization-maintaining fiber,
orthogonally polarized modes propagating along the fast and slow axes experience different propagation constants. As the two counter-propagating beams recombine at the coupler, a cumulative phase difference is introduced between the two orthogonal polarization components.

The resulting phase difference can be expressed as
\begin{equation}
\Delta \phi = \frac{2\pi B L}{\lambda},
\label{eq:pmf_phase}
\end{equation}
where $B$ denotes the birefringence of the PMF, $L$ is the effective length of the PMF segment, and $\lambda$ is the operating wavelength.

As indicated by Eq.~(\ref{eq:pmf_phase}), the phase difference is jointly determined by the birefringence and the effective propagation length of the PMF. Consequently, any external perturbation that modifies $B$ or $L$, such as temperature variation or axial strain, will lead to a measurable shift in the interference spectrum, forming the physical basis for PMF-based Sagnac interferometric sensing.

Accordingly, the free spectral range (FSR) of the interference fringes can be expressed as
\begin{equation}
\Delta \lambda = \frac{\lambda^2}{B L},
\label{eq:pmf_fsr}
\end{equation}
where $\lambda$ is the operating wavelength, $B$ denotes the birefringence of the PMF, and $L$ is the effective length of the PMF segment. As indicated by Eq.~(\ref{eq:pmf_fsr}), the fringe period is inversely proportional to both the fiber birefringence and the propagation length. Therefore, by appropriately selecting the PMF length and birefringence parameters, the spectral fringe spacing can be precisely tailored to ensure sufficient spectral resolvability
within the operating wavelength window.

When the ambient temperature varies, both the birefringence $B$ and the physical length $L$ of the PMF are modified due to the thermo optic effect and thermal expansion, respectively. In general, the birefringence and length of the PMF can be regarded as functions of temperature $T$ and axial strain $\varepsilon$. Their total differentials can therefore be written as
\begin{equation}
\begin{cases}
\mathrm{d}B =
\left(\dfrac{\partial B}{\partial T}\right)\mathrm{d}T
+
\left(\dfrac{\partial B}{\partial \varepsilon}\right)\mathrm{d}\varepsilon, \\[6pt]
\mathrm{d}L =
L\left(\alpha\,\mathrm{d}T + \mathrm{d}\varepsilon\right),
\end{cases}
\label{eq:pmf_differentials}
\end{equation}
where $\varepsilon$ denotes the axial strain applied along the fiber length and $\alpha$ is the linear thermal expansion coefficient of the PMF.

The temperature dependent term $\partial B / \partial T$ originates primarily from the thermo optic effect and temperature induced redistribution of internal stress, whereas the strain dependent term $\partial B / \partial \varepsilon$
arises from the photoelastic effect associated with axial deformation. These coupled dependencies constitute the fundamental mechanism governing the cross sensitive response of the PMF based Sagnac interferometer to temperature and strain perturbations.

To quantitatively describe the influence of external perturbations on the interference spectrum, the phase difference $\Delta \phi$ can be regarded as a function of the birefringence $B$, the effective fiber length $L$, and the operating wavelength $\lambda$. Taking the total differential of
$\Delta \phi = 2\pi B L / \lambda$, the variation of the phase difference can be
expressed as
\begin{equation}
\mathrm{d}(\Delta \phi)
= \frac{2\pi}{\lambda}
\left(
L\,\mathrm{d}B + B\,\mathrm{d}L
\right)
- \frac{2\pi B L}{\lambda^2}\,\mathrm{d}\lambda.
\label{eq:pmf_phase_diff}
\end{equation}

Equation~(\ref{eq:pmf_phase_diff}) establishes the fundamental relationship between the phase variation and the perturbation induced changes in birefringence, fiber length, and wavelength. By substituting the differential expressions of $B$ and $L$ given in Eq.~(\ref{eq:pmf_differentials}) into Eq.~(\ref{eq:pmf_phase_diff}), the wavelength shift of the interference fringes under external perturbations can be explicitly derived.

Under the present experimental conditions, the PMF segment in the Sagnac loop is rigidly fixed and free of external mechanical loading. Consequently, the strain induced contribution is negligible compared with the temperature induced effect, and the wavelength shift of the interference fringes is therefore predominantly governed by temperature variations. As a result, the PMF based Sagnac interferometer exhibits a stable, predictable, and highly repeatable
spectral response to temperature changes, making it well suited as a
temperature dominant reference sensing unit in the cascaded sensing system.

During fabrication, the PMF length was carefully optimized to satisfy the required fringe period and spectral resolvability, as dictated by
Eq.~(\ref{eq:pmf_fsr}). Particular care was taken during fusion splicing to accurately align the fiber axes and polarization directions, thereby maintaining high interference contrast and long term stability. All splicing points were completed under low loss conditions, and repeated spectral scans were conducted to verify the robustness and repeatability of the interferometric response.

\subsection{Preparation of SMS Fiber Structure}

To construct a reference sensing unit that is sensitive to temperature variations
while remaining insensitive to changes in the external refractive index, a
single mode multimode single mode (SMS) fiber structure was employed in this
work. The SMS structure operates based on multimode interference and is
characterized by a simple configuration, high spectral stability, and strong
immunity to perturbations of the surrounding refractive index.

When light from a single mode fiber is coupled into the multimode fiber segment,
multiple guided modes are simultaneously excited due to mode field mismatch.
These modes propagate along the multimode fiber with different propagation
constants and gradually accumulate relative phase differences. Upon
recombination into the output single mode fiber, coherent superposition of the
excited modes produces a periodic interference pattern in the transmission
spectrum.

For a representative pair of dominant interfering modes, the accumulated phase
difference can be approximated as
\begin{equation}
\Delta \phi = (\beta_m - \beta_n)\,L,
\label{eq:sms_phase}
\end{equation}
where $\beta_m$ and $\beta_n$ denote the effective propagation constants of the
$m$th and $n$th guided modes in the multimode fiber, respectively, and $L$ is the
effective length of the multimode fiber.

Based on Eq.~(\ref{eq:sms_phase}), the free spectral range of the SMS interference
fringes can be approximated as
\begin{equation}
\Delta \lambda \approx \frac{\lambda^2}{\Delta n_{\mathrm{eff}}\,L},
\label{eq:sms_fsr}
\end{equation}
where $\Delta n_{\mathrm{eff}} = n_{\mathrm{eff},m} - n_{\mathrm{eff},n}$
represents the effective refractive index difference between the dominant
interfering modes. Equation~(\ref{eq:sms_fsr}) indicates that the fringe period
can be tailored by appropriately selecting the multimode fiber length and mode
excitation conditions, thereby ensuring sufficient spectral resolvability within
the operating wavelength window.

When the ambient temperature varies, both the refractive index and the physical
length of the multimode fiber are modified due to the thermo optic effect and
thermal expansion, respectively. As a result, the effective propagation
constants and the modal index difference $\Delta n_{\mathrm{eff}}$ become
temperature dependent. To quantitatively describe this temperature induced
response, the differentials of the governing parameters can be expressed as
\begin{equation}
\begin{cases}
\mathrm{d}(\Delta n_{\mathrm{eff}})
=
\left(\dfrac{\partial \Delta n_{\mathrm{eff}}}{\partial T}\right)\mathrm{d}T, \\[6pt]
\mathrm{d}L = L\,\alpha\,\mathrm{d}T,
\end{cases}
\label{eq:sms_differentials}
\end{equation}
where $\alpha$ denotes the linear thermal expansion coefficient of the multimode
fiber, and $\partial \Delta n_{\mathrm{eff}} / \partial T$ characterizes the
effective thermo optic response of the interfering modes.

By differentiating the phase matching condition given in
Eq.~(\ref{eq:sms_phase}) and substituting Eq.~(\ref{eq:sms_differentials}), the
wavelength shift of the SMS interference fringes induced by temperature
variation can be derived as
\begin{equation}
\Delta \lambda
=
\lambda
\left[
\frac{1}{\Delta n_{\mathrm{eff}}}
\left(\frac{\partial \Delta n_{\mathrm{eff}}}{\partial T}\right)
+
\alpha
\right]
\Delta T.
\label{eq:sms_temp_sensitivity}
\end{equation}
Equation~(\ref{eq:sms_temp_sensitivity}) indicates that the temperature
sensitivity of the SMS structure originates from the combined contributions of
the thermo optic effect and thermal expansion.

In contrast, during the experiments the cladding of the multimode fiber remained
intact, and the optical field was predominantly confined within the fiber
structure. Consequently, variations in the surrounding refractive index only
weakly perturb the evanescent field outside the cladding and do not induce
observable shifts in the interference spectrum, rendering the SMS structure
largely insensitive to external refractive index changes.

During fabrication, the single mode and multimode fibers were connected using a
controlled lateral offset fusion splicing technique. By introducing an
appropriate transverse offset, the number of excited modes and their energy
distribution within the multimode fiber can be effectively adjusted, thereby
enhancing the contrast and repeatability of the interference spectrum.
Considering structural simplicity, robustness, and sensing reliability, the SMS
structure was ultimately selected as the temperature reference sensing unit in
the cascaded LPFG sensing system.

\section{Results and Discussion}
\subsection{Configuration optimization of cascaded PMF and LPFG and their spectral response characteristics}

Both the polarization-maintaining fiber (PMF) and the long-period fiber grating (LPFG) have individually demonstrated good sensitivities to temperature and axial strain, as verified in the characterization experiments shown in Fig.~S2 and Fig.~S3, indicating their potential for cascaded multi-parameter sensing applications.
Fig.~1(a--c) illustrate the system configuration and corresponding spectral characteristics when an LPFG and a PMF are simultaneously introduced into a Sagnac interferometer. As shown in Fig.~1(a), the LPFG and PMF are cascaded and placed together within the Sagnac loop, with the intention of enabling multi-parameter sensing through their combined response. However, experimental results reveal that, under this configuration, spectral features originating from different physical mechanisms cannot be effectively distinguished. As shown in Fig.~1(b), the resonance trough of the LPFG is strongly superimposed with the interference fringes of the PMF-based Sagnac interferometer, resulting in severe spectral overlap and ambiguity in feature identification. Further analysis indicates that mode-field mismatch and energy redistribution at the LPFG--PMF junction lead to pronounced over-coupling effects. As illustrated in Fig.~1(c), such over-coupling causes distortion and splitting of spectral troughs, significantly degrading the stability of both the resonance wavelengths and the interference fringes, thereby limiting the applicability of this configuration for high-precision multi-parameter sensing.

\begin{figure}[t]
  \centering
  \includegraphics[width=0.8\linewidth]{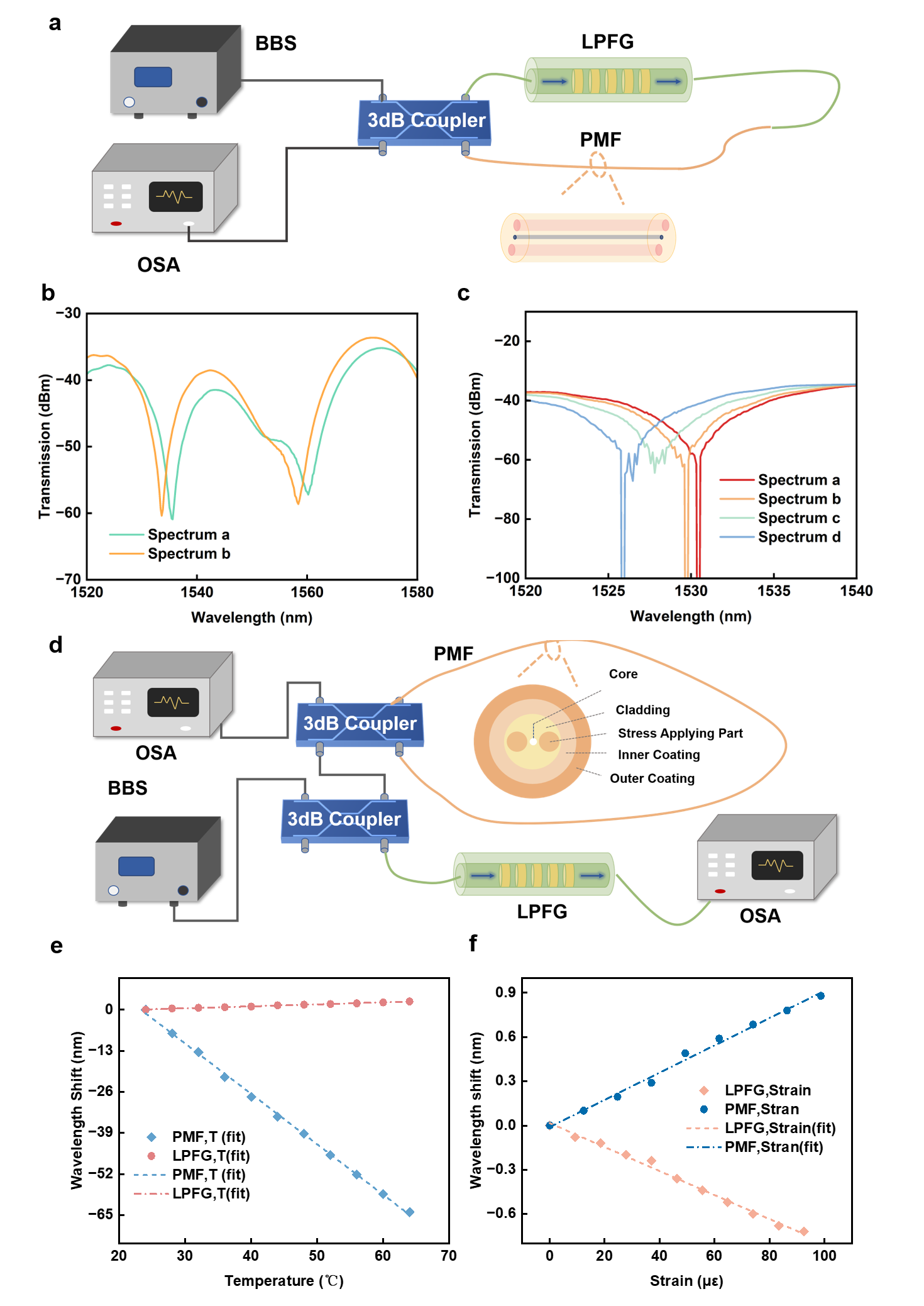}
  \caption{Optical path configuration and spectral responses of the cascaded LPFG--PMF sensing system. (a) Schematic diagram of a Sagnac interferometer incorporating both a long period fiber grating (LPFG) and a polarization-maintaining fiber (PMF) within the same loop. (b) Measured output spectrum of the LPFG--PMF composite Sagnac configuration, where the LPFG resonance trough spectrally overlap with the Sagnac interference fringes. (c) Enlarged view of the coupling region between the LPFG and PMF, illustrating spectral distortion and resonance splitting induced by local over-coupling effects. (d) Optimized sensing configuration, in which the PMF solely forms the Sagnac interferometric loop, while the LPFG operates in-line in transmission along the main optical path. (e) Representative spectral responses of the optimized system under external perturbations, together with the corresponding shifts of selected characteristic wavelengths. (f) Linear dependence of the characteristic wavelength shifts on the applied external physical parameters.
}
  \label{fig:fig1}
\end{figure}

Figure~1(d) presents the improved system configuration, in which the PMF is exclusively employed to form the Sagnac interferometer, while the LPFG is directly cascaded in the main optical path and operates in transmission mode. Under this arrangement, the Sagnac interference spectrum and the LPFG resonance spectrum are effectively separated in both spectral morphology and physical origin, providing a solid foundation for subsequent multi-parameter decoupling and independent demodulation. As shown in Fig.~1(e) and (f), the system exhibits stable and repeatable spectral responses under external perturbations. The LPFG resonance wavelength and the characteristic wavelength of the Sagnac interference spectrum are selected as demodulation parameters, and their dependence on external physical quantities is analyzed through linear fitting.

The results indicate that, within the tested ranges, the wavelength shifts of both characteristic features exhibit excellent linear relationships with the applied physical parameters. For temperature sensing, the PMF-based Sagnac interferometer demonstrates a high sensitivity of approximately $-1.59~\mathrm{nm/^{\circ}C}$, while the LPFG resonance wavelength shows a moderate temperature sensitivity in the range of $0.062$--$0.074~\mathrm{nm/^{\circ}C}$. The corresponding coefficients of determination are all higher than $R^{2} > 0.995$, confirming the excellent linearity of the sensing responses. For strain measurements, the characteristic wavelengths of the Sagnac interferometer and the LPFG exhibit sensitivities of approximately $0.009~\mathrm{nm/\mu\varepsilon}$ and $-0.008~\mathrm{nm/\mu\varepsilon}$, respectively, both maintaining strong linear correlations ($R^{2} > 0.99$).

Based on the linear fitting results, the sensing performance is further evaluated
in terms of sensitivity, linearity, and minimum resolvable quantity. The sensing
resolution is estimated using the three sigma criterion, which can be expressed
as
\begin{equation}
\mathrm{Resolution} = \frac{3\sigma}{S},
\label{eq:resolution}
\end{equation}
where $\sigma$ denotes the standard deviation of the measured characteristic
wavelength under stable environmental conditions, reflecting the intrinsic
wavelength fluctuation of the sensing system, and $S$ represents the
corresponding wavelength sensitivity to the target physical parameter, obtained
from linear regression analysis.

According to Eq.~(\ref{eq:resolution}), a smaller wavelength fluctuation or a
higher sensitivity directly leads to an improved sensing resolution. In the
present experiments, the value of $\sigma$ was experimentally determined from
time series spectral measurements under constant conditions, while the
sensitivity $S$ was extracted from the slope of the fitted wavelength shift
curves.

In addition to resolution analysis, the long term stability of the sensing
system was evaluated by continuously monitoring the output spectra under fixed
environmental conditions. Experimental results indicate that, over a monitoring
duration of $2~\mathrm{h}$, the wavelength drift of both the LPFG resonance
trough and the Sagnac interference feature remains confined within
$\pm 0.02~\mathrm{nm}$. This small fluctuation range demonstrates the excellent
stability and repeatability of the proposed cascaded sensing system, which is
essential for reliable multi parameter sensing and long term operation.

The sensing resolution was evaluated using the three-sigma ($3\sigma$) criterion. 
Here, $\sigma$ denotes the standard deviation of the measured characteristic wavelength obtained from time-series spectral data under constant environmental conditions, reflecting the intrinsic wavelength fluctuation of the system. 
The sensitivity $S$ was determined from the slope of the linear fitting between the wavelength shift and the applied physical parameter. 
Based on the measured wavelength fluctuations and the corresponding sensitivities, the temperature and strain resolutions were estimated to be on the order of $10^{-2}~^\circ\mathrm{C}$ and several microstrains, respectively.

It should be noted that, despite the clear advantages of this improved configuration in terms of spectral decoupling and measurement accuracy, its implementation typically requires simultaneous monitoring of the LPFG resonance spectrum and the Sagnac interference spectrum, which may involve the use of dual spectral interrogation channels. This inevitably increases system complexity and hardware resource consumption. In addition, since the LPFG and PMF interferometric elements are located in different optical paths, environmental perturbations or optical path mismatches may introduce additional measurement uncertainties under extreme conditions. Therefore, while this configuration is well suited as a high-precision experimental platform, practical deployments may require a careful balance between system integration, resource efficiency, and measurement reliability.

\subsection{Comparison of spectral characteristics for different fiber connection configurations}

Although the configuration discussed in Fig.~1 provides insight into the combined spectral behavior of cascaded sensing elements, its practical implementation remains suboptimal due to the increased system complexity and the requirement for multiple spectral interrogation channels, which leads to inefficient utilization of measurement resources. To clarify the role of fiber connection topology in shaping the spectral response, the transmission characteristics of two representative configurations were systematically compared under identical experimental conditions, as summarized in Fig.~2.

\begin{figure}[t]
  \centering
  \includegraphics[width=0.8\linewidth]{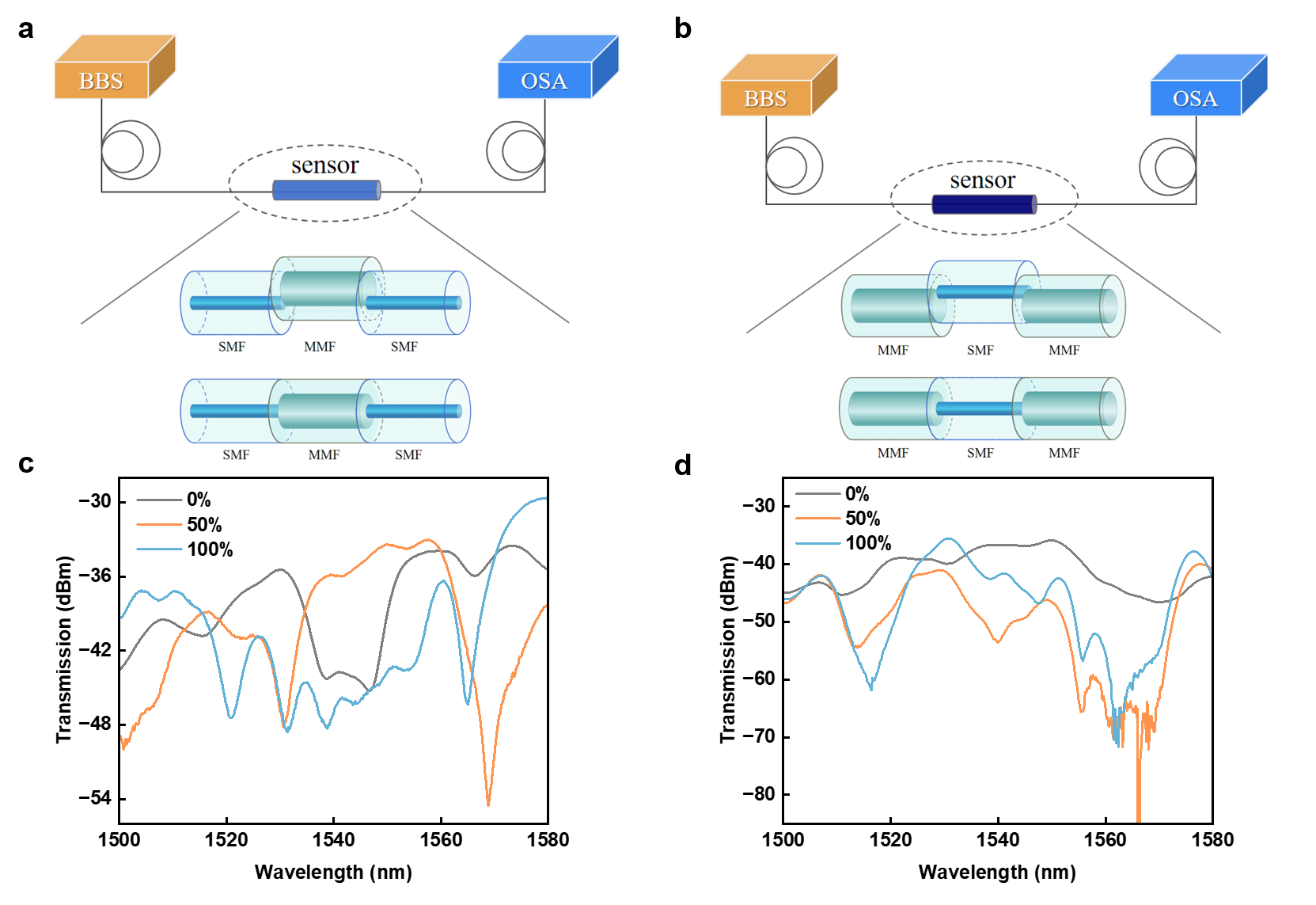}
  \caption{Comparison of spectral responses for different fiber connection configurations.
(a) Schematic illustration of configuration~I, employing an SMF--MMF--SMF structure with a controlled lateral offset at the splicing interfaces to induce multimode interference.
(b) Schematic illustration of configuration~II (MMF--SMF--MMF), where the inserted SMF section acts as a spatial mode filter and largely suppresses effective multimode excitation in the adjacent MMF segment. As a result, the transmitted spectrum is dominated by smooth insertion-loss variations rather than well-defined interference-induced troughs, which limits its suitability for wavelength-tracking-based sensing.
(c) Transmission spectra obtained from configuration~I under different offset conditions, showing that an offset of approximately 50\% yields the most distinct and reproducible resonance features.
(d) Transmission spectra corresponding to configuration~II, exhibiting weak spectral modulation and the absence of well-defined characteristic trough.}
  \label{fig:fig2}
\end{figure}

In configuration~I (Fig.~2(a)), a single-mode--multimode--single-mode (SMF--MMF--SMF) structure was employed, where controlled lateral offsets were deliberately introduced at the SMF--MMF splicing interfaces to promote multimode excitation and interference within the multimode fiber. By varying the offset degree, both the number of excited modes and their relative power distribution can be effectively tailored. Among the tested conditions, an offset of approximately 50\% of the fiber core diameter provides an optimal compromise between coupling efficiency and modal diversity, leading to enhanced and well-balanced multimode interference.

The corresponding transmission spectra are shown in Fig.~2(c). Under the 50\% offset condition, the spectrum exhibits pronounced, stable, and highly reproducible resonance features with well-defined spectral trough. The positions and shapes of the characteristic troughs remain consistent across repeated measurements, indicating robust interference behavior suitable for reliable wavelength tracking and signal demodulation. In contrast, insufficient offset results in weak multimode excitation and reduced spectral modulation, whereas excessive offset introduces significant coupling loss and spectral distortion, both of which degrade the clarity and repeatability of the interference features. These results demonstrate that the lateral offset plays a critical role in determining the quality of the interference spectrum, with the 50\% offset condition yielding the most favorable overall performance.

For comparison, configuration~II is illustrated in Fig.~2(b), in which an SMF section is inserted between two MMF segments (MMF--SMF--MMF). In contrast to the canonical SMS configuration, where the SMF fundamental mode is launched into the MMF to simultaneously excite multiple guided modes that subsequently interfere and form distinct spectral valleys, the intermediate SMF in configuration~II effectively filters the optical field and reduces the modal diversity that can be sustained in the MMF segment. Consequently, the measured spectra (Fig.~2(d)) exhibit weak and unstable spectral modulation, with no distinct, repeatable troughs suitable for robust wavelength demodulation. This behavior is consistent with the established SMS/MMI mechanism, in which clear interference features require sufficient higher-order-mode excitation and balanced modal power distribution, typically achieved through appropriate mode-field mismatch/controlled splicing conditions rather than a mode-filtering topology. Therefore, configuration~II was not adopted for subsequent sensing experiments, and the SMF--MMF--SMF structure with deliberate offset splicing (configuration~I) was selected as the fundamental interference unit.

Overall, the above comparison clearly indicates that the spectral response of the system is strongly governed by the fiber connection configuration. Only configurations that deliberately introduce controlled multimode excitation and interference can generate clear, stable, and analyzable spectral features. Based on these considerations, the SMF--MMF--SMF configuration with an approximately 50\% lateral offset, as shown in Fig.~2(a), was selected as the fundamental sensing structure for all subsequent experiments.

It is worth noting that more sophisticated excitation strategies, including higher-order approaches as exemplified in Fig.~S4, were also experimentally explored. However, these configurations failed to provide further improvement in spectral clarity, stability, or repeatability, and in some cases even introduced additional spectral distortion and sensitivity degradation. Therefore, considering both sensing performance and structural simplicity, the SMF--MMF--SMF configuration with an approximately 50\% lateral offset represents the optimal and most reliable choice among the tested configurations.



\subsection{Comparison of sensing responses based on different spectral demodulation schemes}

After identifying an appropriate fiber connection configuration, the suitability of different spectral features as demodulation parameters is further evaluated from a metrological perspective, including sensitivity, linearity, resolution, and stability.

Figure~3(a) presents the transmission spectra of the multimode fiber under different external perturbations. It can be observed that the overall spectral profile changes with the applied stimulus; however, these variations mainly manifest as fluctuations in transmission intensity rather than distinct and traceable spectral shifts. This observation suggests that demodulation based solely on intensity variation may suffer from amplified measurement uncertainty.

Based on this observation, the transmission intensity variation of the multimode
fiber is first employed as the sensing parameter, and the corresponding
quantitative results are shown in Fig.~3(b). The transmission intensity exhibits
pronounced data dispersion and nonlinear behavior with respect to temperature.
Linear fitting yields a sensitivity of approximately
0.031~dBm/$^\circ$C with a coefficient of determination of $R^2 = 0.88$,
indicating poor fitting consistency. Therefore, intensity-based demodulation is not suitable as a primary sensing
parameter for high-precision measurements.

In contrast, when the wavelength position of the spectral trough is selected as the
characteristic demodulation parameter, the sensing performance is significantly
improved. Figures~3(c) and~3(e) illustrate the systematic wavelength shifts of the
characteristic trough under temperature and axial strain perturbations, respectively.
The corresponding linear fitting results are presented in Figs.~3(d) and~3(f),
revealing stable and nearly linear relationships between the wavelength shift and the applied physical quantities.

For temperature sensing, linear regression of the wavelength shift yields a
sensitivity of \textbf{$-0.079~\mathrm{nm}/^\circ\mathrm{C}$} with a coefficient of
determination of \textbf{$R^2 = 0.997$}, indicating excellent linearity and fitting
consistency. Similarly, under axial strain loading, the extracted wavelength based
sensitivity is \textbf{$-0.694~\mathrm{nm}/m\varepsilon$} with a high coefficient
of determination of \textbf{$R^2 = 0.995$}. These results demonstrate that
wavelength-based demodulation provides markedly improved linearity, robustness,
and repeatability compared with intensity-based schemes, making it more suitable
for high-precision fiber-optic sensing applications.

\begin{figure}[t]
  \centering
  \includegraphics[width=0.8\linewidth]{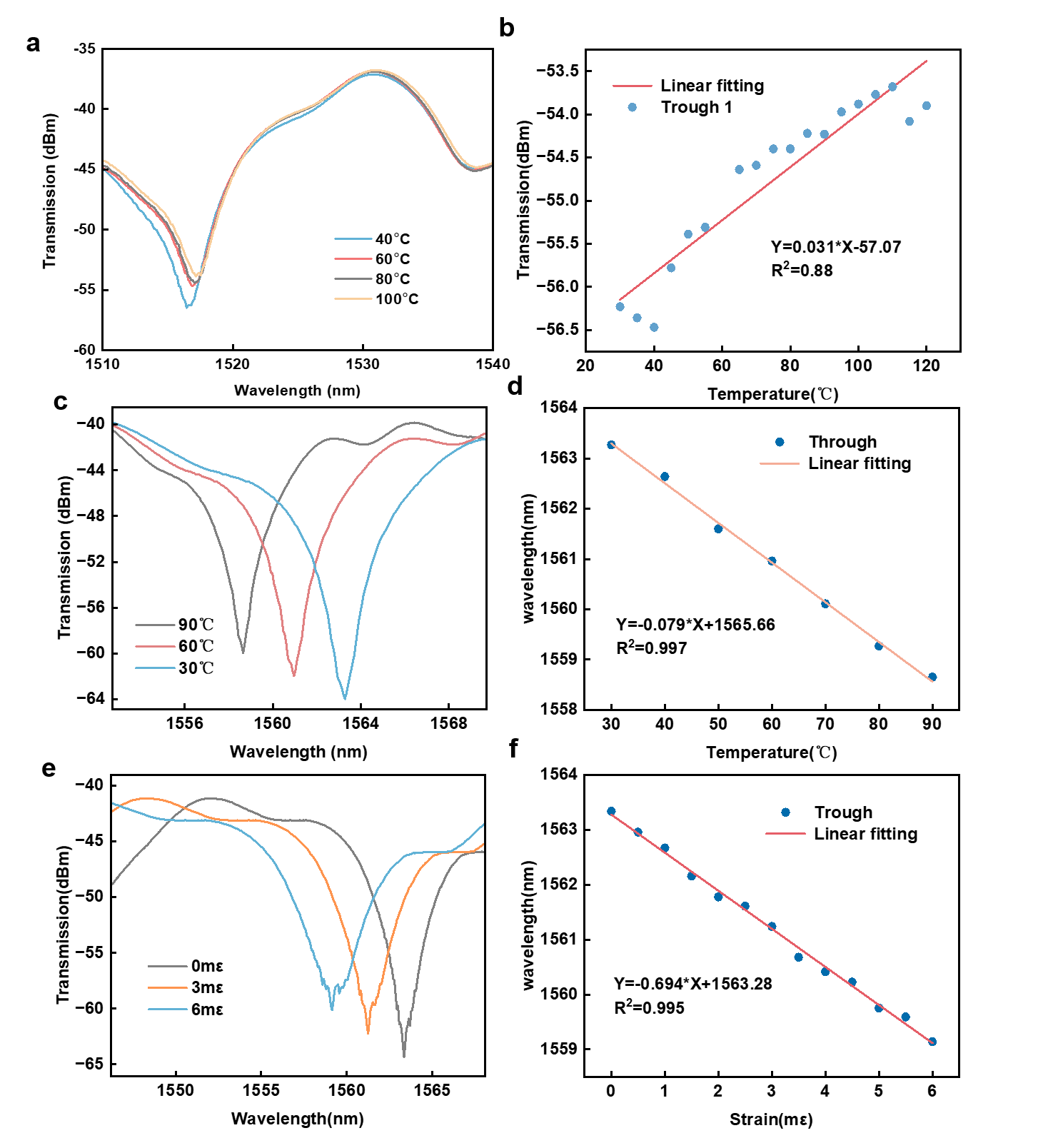}
  \caption{Comparison of sensing responses obtained using different spectral demodulation schemes.
(a) Transmission spectra of the multimode fiber measured at different temperatures (40, 60, 80, and 100~$^\circ$C), where the spectral response is dominated by transmission intensity variation with no well-defined wavelength shift.
(b) Temperature response extracted from intensity-based demodulation of the multimode fiber spectra in (a), exhibiting pronounced data dispersion and limited linearity.
(c) Wavelength shift of a characteristic spectral trough measured at different temperatures (30, 60, and 90~$^\circ$C), demonstrating a clear and monotonic wavelength response.
(d) Linear fitting of the characteristic wavelength as a function of temperature, from which the temperature sensitivity and linearity are extracted.
(e) Wavelength shift of the characteristic spectral trough under different applied axial strain levels (0, 3, and 6~$m\varepsilon$).
(f) Linear fitting of the characteristic wavelength as a function of axial strain, showing improved linearity and repeatability compared with intensity-based demodulation.}
  \label{fig:fig3}
\end{figure}

Furthermore, the minimum resolvable value of the sensing system was evaluated
based on the linear fitting results using the three-sigma criterion where $\sigma$ represents the
standard deviation of the characteristic wavelength fluctuation under a constant
measurand, and $S$ denotes the corresponding wavelength sensitivity. In this
study, $\sigma$ was experimentally determined to be approximately
$0.01~\mathrm{nm}$.

Accordingly, the minimum resolvable temperature change is estimated to be
approximately \textbf{$0.38~^\circ\mathrm{C}$}, while the minimum resolvable axial
strain is approximately \textbf{$0.43~m\varepsilon$}. These results indicate
that the proposed wavelength-based demodulation scheme is capable of resolving
small temperature and strain perturbations with good reliability.

To further assess the short-term stability and repeatability of the sensing system, the characteristic wavelength is continuously monitored while the external conditions are kept constant. The experimental results show that over a monitoring duration of 120~min, the wavelength drift remains within $\pm$ 0.02~nm, demonstrating good temporal stability and measurement repeatability.

Overall, compared with intensity-based demodulation, wavelength-based demodulation exhibits clear advantages in terms of sensitivity, linearity, resolution, and robustness against external disturbances. Consequently, the characteristic wavelength is adopted as the primary demodulation parameter in all subsequent experiments to enable accurate and repeatable temperature and strain measurements.

\subsection{Multi parameter discrimination based on spectral troughs in an SMF--MMF--LPFG cascaded sensor}

\begin{figure}[t]
  \centering
  \includegraphics[width=0.8\linewidth]{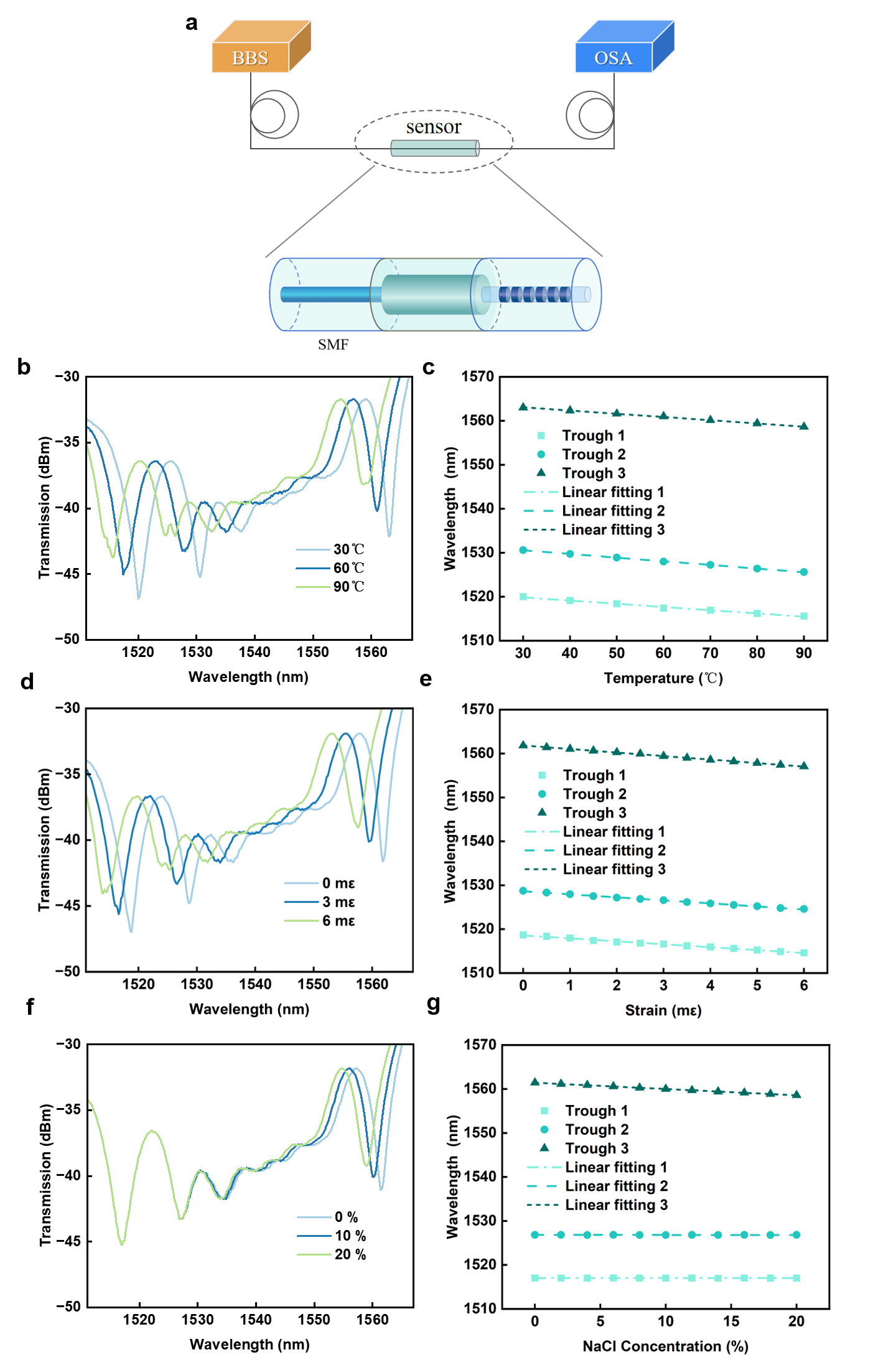}
  \caption{Spectral responses of the SMF--MMF--LPFG cascaded sensor under multiple
external perturbations.
(a) Schematic illustration of the cascaded SMF--MMF--LPFG sensing configuration.
(b) Transmission spectra measured at discrete temperatures of 30, 60, and
90~$^\circ$C, showing multiple characteristic spectral troughs.
(c) Wavelength evolution of three representative spectral troughs as the
temperature increases from 30 to 90~$^\circ$C with a step of 10~$^\circ$C.
(d) Wavelength shifts of the spectral troughs under discrete applied axial strain
levels of 0, 3, and 6~$m\varepsilon$.
(e) Continuous wavelength evolution of the spectral troughs as the axial strain
varies from 0 to 6~$m\varepsilon$ with a step of 1~$m\varepsilon$.
(f) Spectral responses of the characteristic troughs measured at discrete
relative refractive index levels of 0\%, 10\%, and 20\%.
(g) Continuous wavelength evolution of the spectral troughs as the relative
refractive index increases from 0\% to 20\% with a step of 2\%.}
  \label{fig:fig4}
\end{figure}

Based on the optimization of individual sensing units and the wavelength-based
spectral demodulation strategy, a cascaded sensing configuration was constructed
by integrating a SMF, a MMF, and a LPFG within a single optical path, as schematically
illustrated in Fig.~4(a). In this SMF--MMF--LPFG cascaded structure, multiple
spectral troughs originating from different physical mechanisms coexist in the
transmission spectrum. Specifically, the first two characteristic troughs are
generated by multimode interference within the MMF segment, while the third
trough arises from the resonance effect of the LPFG.

It should be emphasized that none of the spectral troughs is exclusively
sensitive to a single physical parameter. Instead, all troughs exhibit
simultaneous responses to temperature, axial strain, and refractive index, but with
distinct sensitivity magnitudes and response trends determined by their
formation mechanisms. Consequently, each spectral trough can be characterized by
a unique sensitivity vector with respect to the target measurands, which forms
the physical basis for multi-parameter discrimination using multi-wavelength
demodulation.

Figures~4(b) and~4(c) present the representative transmission spectra and the
corresponding wavelength evolution of the cascaded sensor under temperature
variation. Both the LPFG resonance trough and the interference-induced troughs
exhibit systematic wavelength shifts with increasing temperature, while their
response magnitudes and trends differ significantly. To quantitatively evaluate
the temperature response, three representative spectral troughs were tracked in
Fig.~4(c). Linear regression analysis shows that trough~1 exhibits a temperature
sensitivity of $-0.0733$~nm/$^\circ$C with a coefficient of determination
$R^2 > 0.992$, trough~2 shows a higher sensitivity of $-0.0833$~nm/$^\circ$C with
$R^2 > 0.996$, and trough~3 presents a sensitivity of $-0.0733$~nm/$^\circ$C with
$R^2 > 0.998$. These results indicate excellent linearity and consistent
temperature-dependent behavior among the multiple spectral troughs.

The sensing responses of the cascaded structure under applied axial strain are
illustrated in Fig.~4(d) and Fig.~4(e). Similar to the temperature case, all
characteristic troughs exhibit clear strain-dependent wavelength shifts.
Quantitative analysis reveals that trough~1 exhibits a strain sensitivity of
$-0.6706$~nm/$m\varepsilon$ with $R^2 > 0.994$, trough~2 shows a sensitivity of
$-0.6916$~nm/$m\varepsilon$ with $R^2 > 0.997$, and trough~3 exhibits a higher
sensitivity of $-0.8026$~nm/$m\varepsilon$ with $R^2 > 0.992$, confirming stable
and repeatable strain responses of the multiple spectral troughs.

To further evaluate the discrimination capability with respect to refractive index, the
wavelength responses of the spectral troughs under refractive index variation were
analyzed, as shown in Fig.~4(f) and Fig.~4(g). It is observed that trough~1 and
trough~2 exhibit extremely weak refractive index sensitivities of less than
$0.0006$~nm/\% and $0.0004$~nm/\%, respectively, indicating that these troughs
remain nearly invariant with refractive index changes. In contrast, trough~3 shows a
pronounced refractive index dependent response with a sensitivity of
$-0.1426$~nm/\% and a coefficient of determination $R^2 > 0.998$. The pronounced
difference in refractive index sensitivity among the spectral troughs further
demonstrates the feasibility of multi-parameter discrimination based on
multi-wavelength spectral demodulation.
Based on the above experimental observations, the wavelength shifts of multiple
spectral troughs can be described in a unified matrix framework. In the general
case, the wavelength shift of the $i$-th spectral trough can be expressed as a
linear superposition of multiple physical parameters,
\begin{equation}
\Delta \lambda_i =
S_{i,T}\,\Delta T
+
S_{i,\varepsilon}\,\Delta \varepsilon
+
S_{i,H}\,\Delta H ,
\label{eq:general_multi_param}
\end{equation}
where $S_{i,T}$, $S_{i,\varepsilon}$, and $S_{i,H}$ denote the temperature, strain,
and refractive index sensitivities of the $i$-th spectral trough, respectively. This
expression indicates that each spectral trough represents a vector response to
multiple measurands rather than an isolated response to a single parameter.

For the specific case of dual parameter decoupling, such as temperature and
axial strain discrimination, two spectral troughs can be selected to construct a
$2\times2$ sensitivity matrix,
\begin{equation}
\begin{bmatrix}
\Delta \lambda_1 \\
\Delta \lambda_2
\end{bmatrix}
=
\begin{bmatrix}
S_{T1} & S_{\varepsilon 1} \\
S_{T2} & S_{\varepsilon 2}
\end{bmatrix}
\begin{bmatrix}
\Delta T \\
\Delta \varepsilon
\end{bmatrix},
\label{eq:dual_param_matrix}
\end{equation}
where the sensitivity coefficients are obtained from independent calibration
experiments. By inverting the sensitivity matrix in
Eq.~(\ref{eq:dual_param_matrix}), the actual temperature and strain variations
can be retrieved directly from the measured wavelength shifts. This formulation
provides a rigorous mathematical representation of the multi wavelength
demodulation strategy and establishes a direct quantitative link between the
measured spectral responses and the target physical parameters.

In addition to multi parameter demodulation capability, long term stability
tests were conducted under fixed environmental conditions to assess the
robustness of the cascaded sensing system. Over a continuous monitoring duration
of $120~\mathrm{min}$, the wavelength fluctuations of all characteristic spectral
troughs were confined within $\pm 0.01~\mathrm{nm}$, indicating excellent
stability and repeatability of the proposed SMF--MMF--LPFG cascaded sensor.

\subsection{Experimental validation under coupled temperature disturbance}

To evaluate the practical performance of the proposed cascaded sensing strategy under realistic cross-sensitive conditions, application-oriented experiments were conducted for both temperature-compensated concentration sensing and temperature--strain decoupled mechanical sensing. In both scenarios, the target measurands were intentionally subjected to pronounced temperature disturbances to emulate field environments where thermal fluctuations are unavoidable.

\begin{figure}[t]
  \centering
  \includegraphics[width=0.8\linewidth]{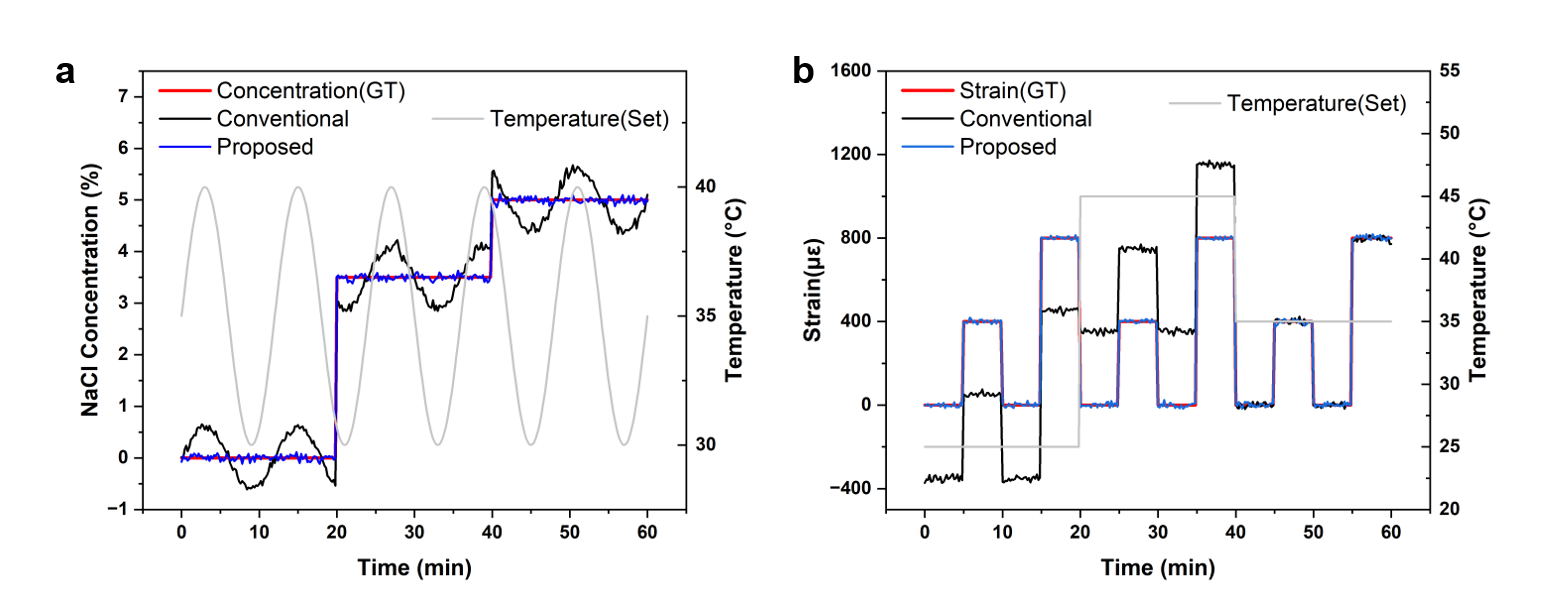}
  \caption{Experimental validation of the proposed multi-parameter decoupling strategy under coupled temperature disturbances.
  (a) Dynamic NaCl concentration tracking under periodic temperature modulation. The concentration ground truth is programmed as stepwise levels of 0\%, 3.5\%, and 5\%, while the temperature is continuously varied between approximately 30~$^\circ$C and 40~$^\circ$C. The conventional single-parameter method exhibits pronounced temperature-induced concentration fluctuations, whereas the proposed method accurately tracks the true concentration with significantly reduced error.
  (b) Temperature--strain decoupled mechanical sensing under concurrent strain loading and temperature steps. The conventional method shows substantial apparent strain drift caused by temperature variations, while the proposed method effectively suppresses thermal cross-sensitivity and faithfully reproduces the strain ground truth.}
  \label{fig:fig4-4}
\end{figure}

Figure~5(a) presents the dynamic NaCl concentration tracking results under a periodically varying temperature environment. The concentration ground truth (GT) was programmed as a stepwise profile of 0\%, 3.5\%, and 5\%, while the ambient temperature was continuously modulated between approximately 30~$^\circ$C and 40~$^\circ$C, as indicated by the gray curve. As expected, the conventional single-parameter demodulation approach exhibits pronounced apparent concentration fluctuations that are strongly correlated with the temperature oscillations. At a fixed concentration level of 3.5\%, the conventional result shows spurious variations of approximately $\pm$0.4--0.6\%, corresponding to an effective temperature-induced cross-sensitivity on the order of 0.08--0.12~\%/$^\circ$C. Such temperature-driven distortions would lead to significant misinterpretation of the actual salinity state in practical applications.

In contrast, the proposed decoupling method accurately tracks the programmed concentration steps and remains largely immune to the imposed temperature disturbance. Across the entire temperature modulation range, the residual concentration fluctuation is confined within $\pm$0.05\%, representing an order-of-magnitude reduction in error compared to the conventional approach. These results demonstrate that the cascaded spectral features combined with matrix-based demodulation enable robust concentration retrieval even under strong thermal perturbations.

The thermo-mechanical decoupling capability of the proposed scheme is further validated in Fig.~5(b), where stepwise strain loading is applied concurrently with abrupt temperature changes. The strain ground truth follows a sequence of 0, 400, and 800~$\mu\varepsilon$, while the temperature is stepped from approximately 25~$^\circ$C to 45~$^\circ$C and subsequently reduced to 35~$^\circ$C. The conventional method exhibits substantial temperature-induced strain artifacts. When the temperature changes by 20~$^\circ$C, the apparent strain offset reaches several hundred microstrain, corresponding to an effective temperature--strain cross-sensitivity of approximately 30--35~$\mu\varepsilon$/$^\circ$C. This level of thermal interference is comparable to or even exceeds the actual mechanical strain being measured.

By contrast, the proposed method successfully suppresses the thermal cross-effect and faithfully reproduces the strain ground truth throughout the entire experiment. Even during rapid temperature transitions, the decoupled strain signal remains stable, with residual deviations limited to approximately $\pm$10--20~$\mu\varepsilon$, which is close to the intrinsic noise floor of the system. Meanwhile, the temperature estimation remains unaffected by the applied mechanical loading, confirming the mutual independence of the retrieved parameters. Overall, these results verify that the proposed cascaded sensing strategy enables reliable multi-parameter discrimination under strongly coupled thermal--chemical and thermo-mechanical disturbances.

\section{Conclusion}

In this work, a cascaded fiber optic sensing strategy was developed to enable
robust multi parameter discrimination under strong cross sensitive conditions.
By systematically optimizing the fiber connection topology and the spectral
demodulation scheme, stable and traceable spectral troughs were generated and
employed as the primary interrogation features for quantitative sensing.

First, the sensing system configuration was optimized to avoid spectral overlap
and feature ambiguity. When the LPFG and PMF were simultaneously placed within a
Sagnac loop, severe superposition between the LPFG resonance troughs and the
PMF induced interference fringes occurred, accompanied by distortion and
splitting caused by over coupling at the junction. An improved configuration was
then adopted, in which the PMF solely formed the Sagnac interferometric loop and
the LPFG operated in line in transmission. This arrangement effectively
separated the spectral signatures originating from different physical mechanisms
and provided a stable foundation for subsequent multi parameter demodulation.

Second, the role of fiber connection topology in shaping the spectral response
was clarified by comparing two representative configurations. The SMF MMF SMF
structure with controlled lateral offset splicing was shown to generate
pronounced, reproducible, and well defined interference troughs, whereas a
straightforward transmission configuration produced only weak modulation without
distinct characteristic features. The SMF MMF SMF configuration with an
approximately 50\% lateral offset was therefore selected as the fundamental
interference unit.

Third, the metrological suitability of different demodulation parameters was
evaluated. Intensity based demodulation exhibited substantial dispersion and
limited linearity with temperature (0.031~dBm/$^\circ$C, $R^2=0.88$), indicating
high susceptibility to source fluctuation, coupling loss variation, and
environmental noise. In contrast, wavelength based demodulation yielded stable
and highly linear responses for both temperature and strain, with sensitivities
of $-0.079$~nm/$^\circ$C ($R^2=0.997$) and $-0.0694$~nm/$m\varepsilon$
($R^2=0.995$), respectively. With a wavelength fluctuation standard deviation of
$\sigma=0.01$~nm, the corresponding minimum resolvable changes were estimated to
be 0.38~$^\circ$C and 0.43~$m\varepsilon$ using the three sigma criterion. A
short term stability test further confirmed good repeatability, with wavelength
drift remaining within $\pm$0.02~nm over 120~min.

Building on these optimizations, an SMF MMF LPFG cascaded sensor was
constructed and demonstrated to support multi parameter discrimination based on
multi wavelength spectral troughs. Multiple characteristic troughs coexisted in
the transmission spectrum, where troughs~1--2 originated from MMF multimode
interference and trough~3 was governed by the LPFG resonance mechanism. None of
the troughs was exclusively sensitive to a single measurand; instead, each
exhibited a distinct sensitivity vector with respect to temperature, axial
strain, and refractive index. Experimentally, the tracked troughs showed excellent
linearity and repeatability under temperature variation (e.g., trough~1:
$-0.0733$~nm/$^\circ$C, $R^2>0.992$; trough~2: $-0.0833$~nm/$^\circ$C, $R^2>0.996$;
trough~3: $-0.0733$~nm/$^\circ$C, $R^2>0.998$) and under axial strain loading
(trough~1: $-0.6706$~nm/$m\varepsilon$, $R^2>0.994$; trough~2:
$-0.6916$~nm/$m\varepsilon$, $R^2>0.997$; trough~3: $-0.8026$~nm/$m\varepsilon$,
$R^2>0.992$). For refractive index, troughs~1--2 were nearly invariant
($<0.0006$~nm/\% and $<0.0004$~nm/\%), while trough~3 exhibited a pronounced
response ($-0.1426$~nm/\%, $R^2>0.998$), confirming the feasibility of
multi parameter discrimination via multi wavelength demodulation. The cascaded
sensor also exhibited good short term stability, with wavelength fluctuations
confined within $\pm$0.01~nm over 120~min under fixed environmental conditions.

Finally, application oriented experiments were conducted to validate the
proposed decoupling strategy under coupled temperature disturbances. In dynamic
NaCl concentration tracking with periodic temperature modulation, the
conventional single parameter method suffered from significant temperature driven
artifacts, whereas the proposed method accurately tracked the ground truth with
substantially reduced error. In thermo mechanical tests with concurrent strain
loading and temperature steps, the conventional approach exhibited large
apparent strain drifts induced by temperature changes, while the proposed method
effectively suppressed thermal cross sensitivity and faithfully reproduced the
strain ground truth, with residual deviations close to the intrinsic noise floor.
These results collectively verify that the proposed cascaded sensing framework,
combined with matrix form multi wavelength demodulation, enables reliable
multi parameter retrieval in environments with unavoidable cross sensitive
coupling.

Future work will focus on extending the sensitivity matrix formulation to fully
coupled three parameter inversion with improved conditioning, enhancing
long term stability through packaging and reference channel compensation, and
exploring compact interrogation hardware for field deployable multi parameter
sensing applications.

\bibliographystyle{IEEEtran}
\bibliography{refs}

\end{document}